\documentclass[aps,pra,twocolumn,showpacs,superscriptaddress]{revtex4}
\usepackage{graphicx}
\usepackage{bm}

\begin{document}

\title{Fractional ac Josephson effect in
  $\bm p$- and $\bm d$-wave superconductors}

\author{Hyok-Jon Kwon}
\affiliation{Department of Physics, University of Maryland, College
  Park, Maryland 20742-4111}

\author{K. Sengupta} 
\affiliation{Department of Physics, Yale University,
   New Haven, Connecticut 06520-8120}

\author{Victor M. Yakovenko}
\affiliation{Department of Physics, University of Maryland, College
  Park, Maryland 20742-4111}

\date{\bf cond-mat/0210148, v.1 8 October 2002, v.5 13 March 2004}


\begin{abstract}
  For certain orientations of Josephson junctions between two
  $p_x$-wave or two $d$-wave superconductors, the subgap Andreev bound
  states produce a $4\pi$-periodic relation between the Josephson
  current $I$ and the phase difference $\phi$: $I\propto\sin(\phi/2)$.
  Consequently, the ac Josephson current has the fractional frequency
  $eV/\hbar$, where $V$ is the dc voltage.  In the tunneling limit,
  the Josephson current is proportional to the first power (not
  square) of the electron tunneling amplitude.  Thus, the Josephson
  current between unconventional superconductors is carried by single
  electrons, rather than by Cooper pairs.  The fractional ac Josephson
  effect can be observed experimentally by measuring frequency
  spectrum of microwave radiation from the junction.  We also study
  junctions between singlet $s$-wave and triplet $p_x$-wave, as well
  as between chiral $p_x\pm ip_y$-wave superconductors.
\end{abstract} 

\maketitle

\section{Introduction}

In many materials, the symmetry of the superconducting order parameter
is unconventional, i.e.\ not $s$-wave.  In the high-$T_c$ cuprates, it
is the singlet $d_{x^2-y^2}$-wave \cite{vanHarlingen}.  There is
experimental evidence that, in the quasi-one-dimensional (Q1D) organic
superconductors $\rm(TMTSF)_2X$ \cite{TMTSF}, the symmetry is triplet
\cite{Chaikin}, most likely the $p_x$-wave \cite{Abrikosov82,Lebed},
with the $x$ axis along the conducting chains.  Experiments indicate
that $\rm Sr_2RuO_4$ has the triplet chiral $p_x\pm ip_y$-wave pairing
symmetry \cite{Maeno}.

The unconventional pairing symmetry typically results in formation of
subgap Andreev bound states \cite{Andreev} on the surfaces of these
superconductors \cite{Bruder}.  For $d$-wave cuprate superconductors,
the midgap Andreev states were predicted theoretically in Ref.\ 
\cite{Hu} and observed experimentally as a zero-bias conductance peak
in tunneling between normal metals and superconductors (see review
\cite{Tanaka-review}).  For the Q1D organic superconductors, the
midgap states were theoretically predicted to exist at the edges
perpendicular to the chains \cite{Ours1,Tanuma}.  In the chiral
superconductor $\rm Sr_2RuO_4$, the subgap surface states are expected
to have a chiral energy dispersion \cite{Honerkamp98}.  Their
contribution to tunneling is more complicated \cite{we-Sr2RuO4} than a
simple zero-bias conductance peak found for the midgap Andreev states.
Various ways of observing electron edge states experimentally are
discussed in Ref.\ \cite{we-Synth}.
 
When two unconventional superconductors are joined together in a
Josephson junction, their Andreev surface states hybridize to form
Andreev bound states in the junction.  These states play an important
role in the Josephson current through the junction \cite{Kulik}.
Andreev bound states in high-$T_c$ junctions were reviewed in Ref.\ 
\cite{Wendin-review}.  The Josephson effect between two Q1D $p_x$-wave
superconductors was studied in Refs.\ \cite{Tanaka-pp,Vaccarella}.
Andreev reflection \cite{Andreev-reflection} at the interfaces between
the A and B phases of superfluid $^3$He was studied in Ref.\ 
\cite{Nishida}.  However, Andreev bound states were not mentioned in
this paper.

In the present paper, we predict a new effect for Josephson junctions
between unconventional nonchiral superconductors, which we call the
fractional ac Josephson effect.  Suppose both superconductors forming
a Josephson junction have surface midgap states originally.  This is
the case for the butt-to-butt junction between two $p_x$-wave Q1D
superconductors, as shown in Fig.\ \ref{fig:JJ}(a), and for the
$45^\circ/45^\circ$ in-plane junction between two $d$-wave
superconductors, as shown in Fig.\ \ref{fig:d-wave}(a).  (The two
angles indicate the orientation of the junction line relative to the
$\bm{b}$ axes of each $d_{x^2-y^2}$ superconductor.)  We predict that
the contribution of the hybridized Andreev bound states produces a
$4\pi$-periodic relation between the supercurrent $I$ and the
superconducting phase difference $\phi$: $I\propto\sin(\phi/2)$
\cite{2phi}.  Consequently, the ac Josephson effect has the frequency
$eV/\hbar$, where $e$ is the electron charge, $V$ is the applied dc
voltage, and $\hbar$ is the Planck constant.  The predicted frequency
is a half of the conventional Josephson frequency $2eV/\hbar$
originating from the conventional Josephson relation
$I\propto\sin{\phi}$ with the period of $2\pi$.  Qualitatively, the
predicted effect can be interpreted as follows.  The Josephson current
across the two unconventional superconductors is carried by tunneling
of \emph{single electrons} (rather than Cooper pairs) between the two
resonant midgap states.  Thus, the Cooper pair charge $2e$ is replaced
the single charge $e$ in the expression for the Josephson frequency.
This interpretation is also supported by the finding that, in the
tunneling limit, the Josephson current is proportional to the first
power (not square) of the electron tunneling amplitude
\cite{Tanaka2,Riedel,Barash}.  Possibilities for experimental
observation of the fractional ac Josephson effect are discussed in
Sec.\ \ref{sec:experiment}.  A summary of this work is published in
the conference proceedings \cite{Brazil}.

The predicted current-phase relation $I\propto\sin(\phi/2)$ is quite
radical, because every textbook on superconductivity says that the
Josephson current must be $2\pi$-periodic in the superconducting phase
difference $\phi$ \cite{2phi}.  To our knowledge, the only paper that
discussed the $4\pi$-periodic Josephson effect is Ref.\ \cite{Kitaev}
by Kitaev.  He considered a highly idealized model of spinless
fermions on a one-dimensional (1D) lattice with superconducting
pairing on the neighboring sites.  The pairing potential in this case
has the $p_x$-wave symmetry, and midgap states exist at the ends of
the chain.  They are described by the Majorana fermions, which Kitaev
proposed to use for nonvolatile memory in quantum computing.  He found
that, when two such superconductors are brought in contact, the system
is $4\pi$-periodic in the phase difference between the
superconductors.  Our results are in agreement with his work.
However, we formulate the problem as an experimentally realistic
Josephson effect between known superconducting materials.

For completeness, we also calculate the spectrum of Andreev bound
states and the Josephson current between a singlet $s$-wave and a
triplet $p$-wave superconductors, as well as between two chiral
$p$-wave superconductors \cite{Barash-chiral}.  In agreement with
previous literature \cite{Yip,Tanaka-sp,Asano}, we find that a
Josephson current is permitted between singlet and triplet
superconductors, contrary to a common misconception that it is
forbidden by the symmetry difference.  However, we do not find the
fractional Josephson effect in these cases.

\section{The basics}

The spin symmetry of the Cooper pairing is classified as either
singlet
$\langle\hat{c}_\sigma(\bm{k})\hat{c}_{\sigma'}(-\bm{k})\rangle
\propto\epsilon_{\sigma\sigma'}\Delta(\bm{k})
=i\hat\sigma^{(y)}_{\sigma\sigma'}\Delta(\bm{k})$ or triplet
$\langle\hat{c}_\sigma(\bm{k})\hat{c}_{\sigma'}(-\bm{k})\rangle
\propto i\hat\sigma^{(y)}(\hat{\bm{\sigma}}\cdot\bm{n})
\Delta(\bm{k})$ \cite{Leggett}.  Here $\hat{c}_\sigma(\bm{k})$ is the
annihilation operator of an electron with the spin $\sigma$ and
momentum $\bm{k}$; $\epsilon_{\sigma\sigma'}$ is the antisymmetric
metric tensor and $\hat{\mbox{\boldmath$\sigma$}}$ are the Pauli
matrices acting in the spin space; $\bm{n}$ is a unit vector
characterizing polarization of the triplet state.  In this paper, we
consider only the class of triplet superconductors where the
spin-polarization vector $\bm{n}$ has a uniform, momentum-independent
orientation.  Everywhere in the paper, except in Sec.\ 
\ref{sec:nonparallel}, we select the spin quantization axis $z$ along
the vector $\bm{n}$.  Then the Cooper pairing takes place between
electrons with the opposite $z$-axis spin projections $\sigma$ and
$\bar\sigma$: $\langle\hat c_\sigma(\bm{k})\hat
c_{\bar\sigma}(-\bm{k})\rangle \propto\Delta_\sigma(\bm{k})$.  Because
the fermion operators $\hat c$ anticommute, the pairing potential has
the symmetry $\Delta_\sigma(\bm{k})=\mp\Delta_{\bar\sigma}(\bm{k})
=\pm\Delta_\sigma(-\bm{k})$, where the upper and lower signs
correspond to the singlet and triplet cases.

We select the coordinate axis $x$ perpendicular to the Josephson
junction plane.  We assume that the interface between the two
superconductors is smooth enough, so that the electron momentum
component $k_y$, parallel to the junction plane, is a conserved good
quantum number.

Electron states in a superconductor are described by the Bogoliubov
operators $\hat\gamma$, which are related to the electron operators
$\hat c$ by the following equations \cite{Zagoskin}
\begin{eqnarray}
  && \hat{\gamma}_{n\sigma k_y} = \int dx\,
  [u_{n\sigma k_y}^*(x) \, \hat{c}_{\sigma k_y}(x)
  +v_{n\sigma k_y}^*(x) \, \hat{c}_{\bar\sigma\bar k_y}^\dag(x)],
\label{gamma_n} \\
  && \hat c_{\sigma k_y}(x) = \sum_{n} 
  [u_{n\sigma k_y}(x) \, \hat\gamma_{n\sigma k_y}  
  + v_{n\bar\sigma\bar k_y}^*(x) \,
  \hat\gamma_{n\bar\sigma\bar k_y}^\dag],
\label{canon}
\end{eqnarray}
where $\bar k_y=-k_y$, and $n$ is the quantum number of the Bogoliubov
eigenstates.  The two-components vectors $\psi_{n\sigma
  k_y}(x)=[u_{n\sigma k_y}(x),v_{n\sigma k_y}(x)]$ are the eigenstates
of the Bogoliubov-de Gennes (BdG) equation with the eigenenergies
$E_{n\sigma k_y}$
\begin{equation}
  \left(\begin{array}{cc} 
  \varepsilon_{k_y}(\hat k_x)+U(x) & 
  \hat\Delta_{\sigma k_y}(x,\hat k_x) \\ 
  \hat\Delta_{\sigma k_y}^\dag(x,\hat k_x) & 
  -\varepsilon_{k_y}(\hat k_x)-U(x)
  \end{array}\right) 
  \psi_{n} = E_{n} \psi_{n},
\label{eq:BdG}
\end{equation}
where $\hat k_x=-i\partial_x$ is the $x$ component of the electron
momentum operator, and $U(x)$ is a potential.  In Eq.\ (\ref{eq:BdG})
and below, we often omit the indices $\sigma$ and $k_y$ to shorten
notation where it does not cause confusion.

\section{Junctions between quasi-one-dimensional superconductors}
\label{sec:Q1D}

In this section, we consider junctions between two Q1D
superconductors, such as organic superconductors $\rm(TMTSF)_2X$, with
the chains along the $x$ axis, as shown in Fig.\ \ref{fig:JJ}(a).  For
a Q1D conductor, the electron energy dispersion in Eq.\ (\ref{eq:BdG})
can be written as $\varepsilon=\hbar^2\hat
k_x^2/2m-2t_b\cos(bk_y)-\mu$, where $m$ is an effective mass, $\mu$ is
the chemical potential, $b$ and $t_b$ are the distance and the
tunneling amplitude between the chains.  The superconducting pairing
potentials in the $s$- and $p_x$-wave cases have the forms
\begin{eqnarray}
  \hat\Delta_{\sigma k_y}(x,\hat k_x) &=& \left\{ 
  \begin{array}{cc}
  \sigma\Delta_\beta, & \mbox{$s$-wave}, \\
  \Delta_\beta\,\hat k_x/k_F , & \mbox{$p_x$-wave},
  \end{array} \right.
\label{hat-Delta}
\end{eqnarray}
where $\hbar k_F=\sqrt{2m\mu}$ is the Fermi momentum, and $\sigma$ is
treated as $+$ for $\uparrow$ and $-$ for $\downarrow$.  The index
$\beta=R,L$ labels the right ($x>0$) and left ($x<0$) sides of the
junction, and $\Delta_\beta$ acquires a phase difference $\phi$ across
the junction:
\begin{equation}
  \Delta_R=\Delta_0e^{i\phi}, \qquad
  \Delta_L=\Delta_0~.
\label{gapfn}
\end{equation}
The potential $U(x)=U_0\delta(x)$ in Eq.\ (\ref{eq:BdG}) represents
the junction barrier located at $x=0$.  Integrating Eq.\ 
(\ref{eq:BdG}) over $x$ from --0 to +0, we find the boundary
conditions at $x=0$:
\begin{eqnarray}
  && \psi_L=\psi_R,\quad
  \partial_x\psi_R - \partial_x\psi_L = k_F Z\,\psi(0),
\label{Bdy} \\
  && Z=2mU_0/\hbar^2k_F, \quad D=4/(Z^2+4),
\label{ZDG}
\end{eqnarray}
where $D$ is the transmission coefficient of the barrier.

\subsection{Andreev bound states}
\label{sec:bound}

A general solution of Eq.\ (\ref{eq:BdG}) is a superposition of the
terms with the momenta close to $\alpha k_F$, where the index
$\alpha=\pm$ labels the right- and left-moving electrons:
\begin{equation}
  \psi_{\beta\sigma} = {e^{\beta\kappa x}} \left[ 
  A_\beta \left(
  \begin{array}{c} u_{\beta\sigma+} \\ v_{\beta\sigma+} \end{array}
  \right) e^{i\tilde k_Fx} 
  + B_\beta \left( 
  \begin{array}{c} u_{\beta\sigma-} \\ v_{\beta\sigma-} \end{array} 
  \right) e^{-i\tilde k_Fx} \right],
\label{L0s}
\end{equation}
where $\beta=\mp$ for $R$ and $L$.  Eq.\ (\ref{L0s}) describes a
subgap state with an energy $|E|<\Delta_0$, which is localized at the
junction and decays exponentially in $x$ within the length $1/\kappa$.
The coefficients $(u_{\beta\sigma\alpha},v_{\beta\sigma\alpha})$ in
Eq.\ (\ref{L0s}) are determined by substituting the right- and
left-moving terms separately into Eq.\ (\ref{eq:BdG}) for $x\neq0$,
where $U(x)=0$.  In the limit $k_F\gg\kappa$, we find
\begin{equation}
  \eta_{\beta\sigma\alpha}
  ={v_{\beta\sigma\alpha} \over u_{\beta\sigma\alpha}}=
  {{E+i\alpha\beta\hbar\kappa v_F} \over \Delta_{\beta\sigma\alpha}},
  \quad  \kappa={\sqrt{\Delta_0^2-|E|^2} \over \hbar v_F}, 
\label{kappa-eta}
\end{equation}
where $v_F=\hbar k_F/m$ is the Fermi velocity, and
\begin{eqnarray}
  \Delta_{\beta\sigma\alpha} &=& \left\{ 
  \begin{array}{cc}
  \sigma\Delta_\beta, & \mbox{$s$-wave}, \\
  \alpha\Delta_\beta, & \mbox{$p_x$-wave},
  \end{array} \right.
\label{Delta_bsa}
\end{eqnarray}
with $\Delta_\beta$ given by Eq.\ (\ref{gapfn}).  The $k_y$-dependent
Fermi momentum $\hbar\tilde k_F=\hbar k_F+2t_b\cos(bk_y)/v_F$ in Eq.\ 
(\ref{L0s}) eliminates the dispersion in $k_y$ from the BdG equation.

Substituting Eq.\ (\ref{L0s}) into the boundary conditions
(\ref{Bdy}), we obtain four linear homogeneous equations for the
coefficients $A_\beta$ and $B_\beta$.  These equations are compatible
if the determinant of the corresponding $4\times4$ matrix is zero.
This compatibility condition has the following form:
\begin{eqnarray}
  && {
  (u_{+\sigma -}v_{-\sigma -}-v_{+\sigma -}u_{-\sigma -})
    (u_{+\sigma +}v_{-\sigma +}-v_{+\sigma +}u_{-\sigma +})  
  \over 
  (u_{+\sigma -}v_{-\sigma +}-v_{+\sigma -} u_{-\sigma +})
    (u_{+\sigma +}v_{-\sigma -}-v_{+\sigma +}u_{-\sigma -}) 
  } \nonumber \\
  &&  = 1-D. 
\label{self-uv}
\end{eqnarray}
Using the variables $\eta$ defined in Eq.\ (\ref{kappa-eta}), Eq.\ 
(\ref{self-uv}) can be written in a simpler form
\begin{equation}
  {
    (\eta_{-\sigma -}-\eta_{+\sigma -})
    (\eta_{-\sigma +}-\eta_{+\sigma +}) 
  \over 
    (\eta_{-\sigma +}-\eta_{+\sigma -})
    (\eta_{-\sigma -}-\eta_{+\sigma +}) 
  } = 1-D. 
\label{self}
\end{equation}
Substituting Eq.\ (\ref{kappa-eta}) into Eq.\ (\ref{self}), we obtain
an equation for the energies of the Andreev bound states.  For a given
$\sigma$, there are two subgap states with the energies
$E_a=aE_0(\phi)$ labeled by the index $a=\pm$, where
\begin{eqnarray}
  E_0^{(s)}(\phi) &=&- \Delta_0 \sqrt{1-D\sin^2(\phi/2)},
  \; \mbox{$s$-$s$ junction},
\label{E_s} \\
  E_0^{(p)}(\phi) &=& -\Delta_0 \sqrt{D}\cos(\phi/2),
  \quad \mbox{$p_x$-$p_x$ junction}.
\label{E_p}
\end{eqnarray}

\begin{figure}[t] 
\includegraphics[width=\linewidth,angle=0]{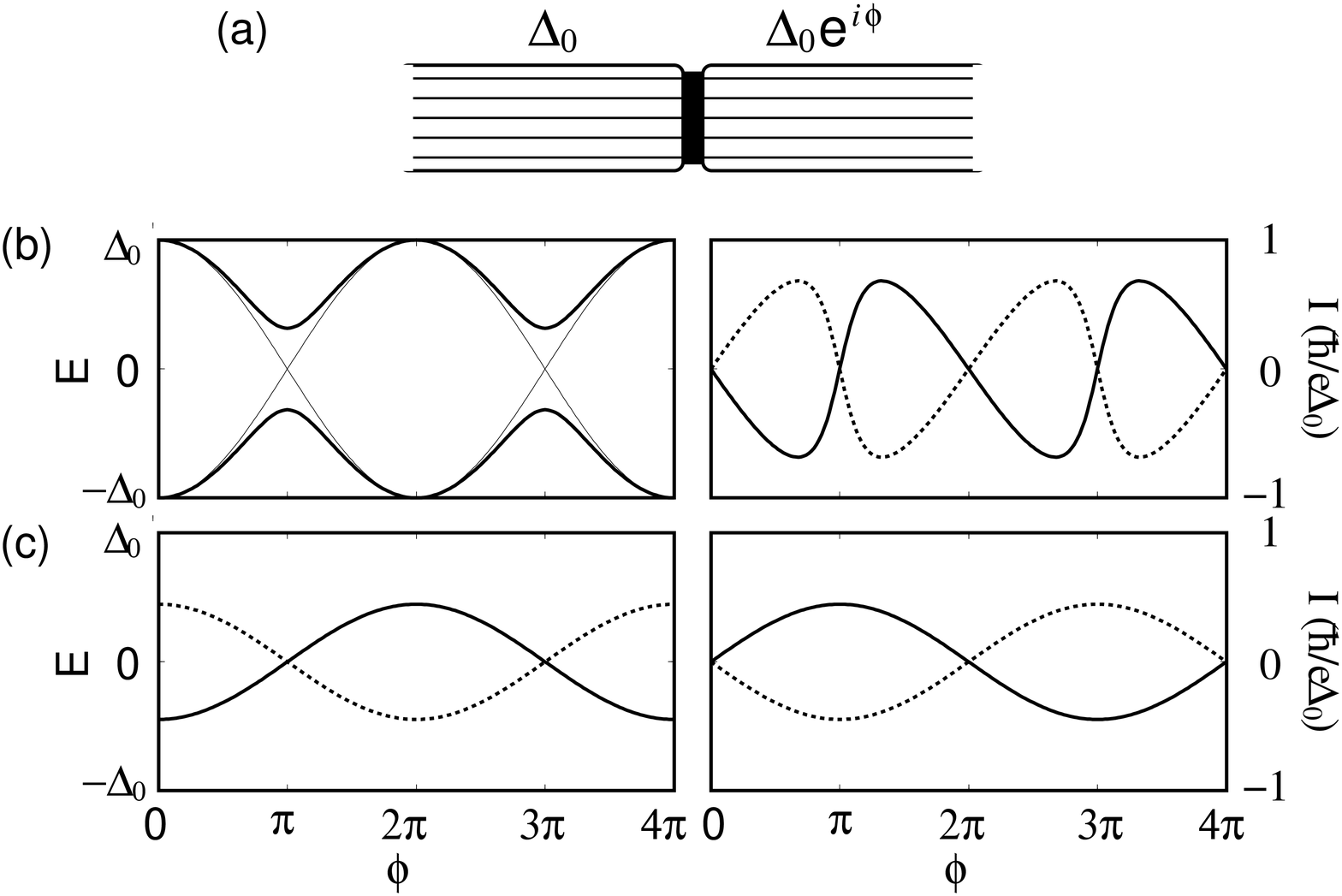} 
\caption{
  (a) Butt-to-butt Josephson junction between two Q1D $p_x$-wave
  superconductors.  (b) The energies (left panel) and the currents
  (right panel) of the Andreev subgap states in the $s$-$s$ junction
  as functions of the phase difference $\phi$ for $D=1$ (thin lines)
  and $D=0.9$ (thick lines).  (c) The same as (b) for the $p_x$-$p_x$
  junction at $D=0.2$.}
\label{fig:JJ}
\end{figure}

The energies (\ref{E_s}) and (\ref{E_p}) are plotted as functions of
$\phi$ in the left panels (b) and (c) of Fig.\ \ref{fig:JJ}.  Without
barrier ($D=1$), the spectra of the $s$-$s$ and $p_x$-$p_x$ junctions
are the same and consist of two crossing curves $E=\mp
\Delta_0\cos\phi/2$, shown by the thin lines in the left panel of
Fig.\ \ref{fig:JJ}(b).  A nonzero barrier ($D<1$) affects the energies
of the Andreev bound states in the $s$-$s$ and $p_x$-$p_x$ junctions
in different ways.  In the $s$-$s$ case, the two energy levels repel
near $\phi=\pi$ and form two separated $2\pi$-periodic branches shown
by the thick lines in the left panel of Fig.\ \ref{fig:JJ}(b).  This
is well known for the $s$-$s$ junctions \cite{Zagoskin,Furusaki99}.
In contrast, in the $p_x$-$p_x$ case, the two energy levels continue
to cross at $\phi=\pi$, and they detach from the continuum of states
above $+\Delta_0$ and below $-\Delta_0$ at $\phi=0$ and $2\pi$, as
shown in the left panel of Fig.\ \ref{fig:JJ}(c).  The absence of
energy levels repulsion at $\phi=\pi$ indicates that there is no
matrix element between these levels in the $p_x$-$p_x$ case, unlike in
the $s$-$s$ case.

As shown in Sec.\ \ref{sec:d-wave}, the $45^\circ/45^\circ$ junction
between two $d$-wave superconductors is mathematically equivalent to
the $p_x$-$p_x$ junction.  Eq.\ (\ref{E_p}) was derived for the
$45^\circ/45^\circ$ junction in Refs.\ \cite{Tanaka1,Riedel,Barash}.

\subsection{dc Josephson effect in thermodynamic equilibrium}
\label{sec:dc}

It is well known \cite{Zagoskin,A-B} that the current carried by a
quasiparticle state $a$ is
\begin{equation}
  I_a={2e \over \hbar}\,{\partial E_a \over \partial\phi}.
\label{eq:I_a}  
\end{equation}
The two subgap states carry opposite currents, which are plotted vs.\ 
$\phi$ in the right panels (b) and (c) of Fig.\ \ref{fig:JJ} for the
$s$-$s$ and $p_x$-$p_x$ junctions.  In thermodynamic equilibrium, the
total current is determined by the Fermi occupation numbers $f_a$ of
the states at a temperature $T$:
\begin{equation}
  I= {2e\over\hbar}\sum_{a=\pm}
  {\partial E_a\over\partial\phi}\,f_a=
  -{2e\over\hbar}{\partial E_0\over\partial\phi}
  \tanh\left({E_0\over2T}\right).
\label{thermal}
\end{equation}
For the $s$-$s$ junction, substituting Eq.\ (\ref{E_s}) into Eq.\ 
(\ref{thermal}), we recover the Ambegaokar-Baratoff formula \cite{AB}
in the tunneling limit $D\ll1$
\begin{equation}
  I_s \approx D\sin\phi\,{e\Delta_0\over2\hbar} 
  \tanh\left({\Delta_0\over2T}\right)
  =\sin\phi\,{\pi\Delta_0\over2eR}
  \tanh\left({\Delta_0\over2T}\right)
\label{I_s}
\end{equation}  
and the Kulik-Omelyanchuk formula \cite{KO} in the transparent limit
$D\to1$  
\begin{equation}
I_s \approx \sin\left({\phi\over2}\right)
  {e\Delta_0\over\hbar} 
  \tanh\left({\Delta_0\cos(\phi/2)\over2T}\right).
\label{I_sKO}
\end{equation}  
Taking into account that the total current is proportional to the
number $N$ of conducting channels in the junction (e.g.\ the number of
chains), we have replaced the transmission coefficient $D$ in Eq.\ 
(\ref{I_s}) by the junction resistance $R=h/2Ne^2D$ in the normal
state.

Substituting Eq.\ (\ref{E_p}) into Eq.\ (\ref{thermal}), we find the
Josephson current in the $p_x$-$p_x$ junction in thermodynamic
equilibrium:
\begin{eqnarray}
  && I_p = \sqrt{D}\sin\left({\phi\over2}\right)
  {e\Delta_0\over\hbar} 
  \tanh\left({\Delta_0\sqrt{D}\cos(\phi/2)\over2T}\right)
\nonumber \\
  && = \sin\left({\phi\over2}\right)
  {\pi\Delta_0\over\sqrt{D}eR}
  \tanh\left({\Delta_0\sqrt{D}\cos(\phi/2)\over2T}\right).
\label{I_p}
\end{eqnarray}

\begin{figure}[t] 
\includegraphics[width=0.8\linewidth,angle=0]{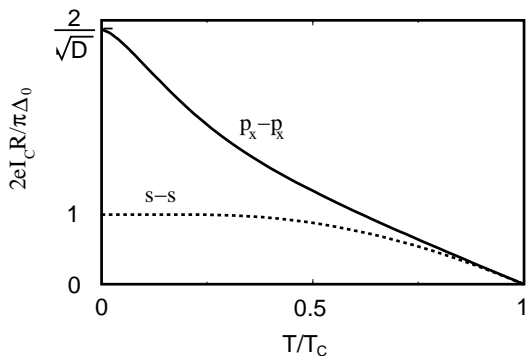} 
\caption{
  Critical currents of the $s$-$s$ (dashed line) and $p_x$-$p_x$
  (solid line) Josephson junctions as functions of temperature for
  $D=0.3$.}
\label{fig:Ic} 
\end{figure}

The temperature dependences of the critical currents for the $s$-$s$
and $p_x$-$p_x$ junctions are shown in Fig.\ \ref{fig:Ic}.  They are
obtained from Eqs.\ (\ref{I_s}) and (\ref{I_p}) assuming the BCS
temperature dependence for $\Delta_0$.  In the vicinity of $T_c$,
$I_p$ and $I_s$ have the same behavior.  With the decrease of
temperature, $I_s$ quickly saturates to a constant value, because, for
$D\ll1$, $E_a^{(s)}\approx\mp\Delta_0$ (\ref{E_s}), thus, for
$T\alt\Delta_0$, the upper subgap state is empty and the lower one is
completely filled.  In contrast, $I_p$ rapidly increases with
decreasing temperature as $1/T$ and saturates to a value enhanced by
the factor $2/\sqrt{D}$ relative to the Ambegaokar-Baratoff formula
(\ref{E_s}) at $T=0$.  This is a consequence of two effects.  As Eqs.\ 
(\ref{I_s}) and (\ref{I_p}) show, $I_s\propto D$ and $I_p\propto
\sqrt{D}$, thus $I_p\gg I_s$ in the tunneling limit $D\ll1$.  At the
same time, the energy splitting between the two subgap states in the
$p_x$-$p_x$ junction is small compared to the gap:
$E_0^{(p)}\propto\sqrt{D}\Delta_0\ll\Delta_0$.  Thus, for
$\sqrt{D}\Delta_0\alt T\alt\Delta_0$, the two subgap states are almost
equally populated, so the critical current has the $1/T$ temperature
dependence analogous to the Curie spin susceptibility.

Eq.\ (\ref{I_p}) was derived analytically for the $45^\circ/45^\circ$
junction between two $d$-wave superconductors in Refs.\ 
\cite{Riedel,Tanaka2}, and a similar result was calculated numerically
for the $p_x$-$p_x$ junction in Ref.\ \cite{Tanaka-pp,Vaccarella}.
Notice that Eq.\ (\ref{I_p}) gives the Josephson current $I_p(\phi)$
that is a $2\pi$-periodic functions of $\phi$, both for $T=0$ and
$T\neq0$.  This is a consequence of the thermodynamic equilibrium
assumption.  At $T=0$, this assumption implies that the subgap state
with the lower energy is occupied, and the one with the higher energy
is empty.  As one can see in Fig.\ \ref{fig:JJ}, the \emph{lower}
energy is always a $2\pi$-periodic functions of $\phi$.  The
assumption of thermodynamic equilibrium was explicitly made in Ref.\ 
\cite{Riedel} and was implicitly invoked in Refs.\ 
\cite{Tanaka2,Tanaka-pp,Vaccarella} by using the Matsubara diagram
technique.  In Ref.\ \cite{Arie}, temperature dependence of the
Josephson critical current was measured in the YBCO ramp-edge
junctions with different crystal angles and was found to be
qualitatively consistent with the upper curve in Fig.\ \ref{fig:Ic}.

\subsection{Dynamical fractional ac Josephson effect}

The calculations of the previous section apply in the static case,
where a given phase difference $\phi$ is maintained for an infinitely
long time, so the occupation numbers of the subgap states have enough
time to relax to thermodynamic equilibrium.  Now let us consider the
opposite, dynamical limit.  Suppose a small voltage $eV\ll\Delta_0$ is
applied to the junction, so the phase difference acquires dependence
on time $t$: $\phi(t)=2eVt/\hbar$.  In this case, the state of the
system is determined dynamically starting from the initial conditions.
Let us consider the $p_x$-$p_x$ junction at $T=0$ in the initial state
$\phi=0$, where the two subgap states (\ref{E_p}) with the energies
$\pm E_0$ are, correspondingly, occupied and empty.  If $\phi(t)$
changes sufficiently slowly (adiabatically), the occupation numbers of
the subgap states do not change.  In other words, the states shown by
the solid and dotted lines in Fig.\ \ref{fig:JJ}(c) remains,
correspondingly, occupied and empty.  The occupied state (\ref{E_p})
produces the current (\ref{eq:I_a}):
\begin{equation}
  I_p(t)
  ={\sqrt{D}e\Delta_0\over\hbar}\sin\left({\phi(t)\over2}\right)
  ={\sqrt{D}e\Delta_0\over\hbar}\sin\left({eVt\over\hbar}\right).
\label{I_t}
\end{equation}
The frequency of the ac current (\ref{I_t}) is $eV/\hbar$, a half of
the conventional Josephson frequency $2eV/\hbar$.  The fractional
frequency can be traced to the fact that the energies Eq.\ (\ref{E_p})
and the corresponding wave functions have the period $4\pi$ in $\phi$,
rather than conventional $2\pi$.  Although at $\phi=2\pi$ the spectrum
in the left panel of Fig.\ \ref{fig:JJ}(c) is the same as at $\phi=0$,
the occupation numbers are different: The lower state is empty and the
upper state is occupied.  Only at $\phi=4\pi$ the occupation numbers
are the same as at $\phi=0$.

The $4\pi$ periodicity is the consequence of the energy levels
crossing at $\phi=\pi$.  (In contrast, in the $s$-wave case, the
levels repel at $\phi=\pi$ in Fig.\ \ref{fig:JJ}(b), thus the energy
curves are $2\pi$-periodic.)  As discussed at the end of Sec.\ 
\ref{sec:bound}, there is no matrix element between the crossing
energy levels at $\phi=\pi$.  Thus, there are no transitions between
them, so the occupation numbers of the solid and dotted curves in
Fig.\ \ref{fig:JJ}(c) are preserved.  In order to show this more
formally, we can write a general solution of the time-dependent BdG
equation as a superposition of the two subgap states with the
time-dependent $\phi(t)$: $\psi(t)=\sum_aC_a(t)\,\psi_a[\phi(t)]$.
The matrix element of transitions between the states is proportional
to $\dot\phi\langle\psi_+|\partial_\phi\psi_-\rangle=\dot\phi
\langle\psi_+|\partial_\phi\hat{H}|\psi_-\rangle/(E_--E_+)$.  We found
that it is zero in the $p_x$-wave case, thus there are no transitions,
and the initial occupation numbers of the subgap states at $\phi=0$
are preserved dynamically.

As one can see in Fig.\ \ref{fig:JJ}(c), the system is not in the
ground state when $\pi<\phi<3\pi$, because the upper energy level is
occupied and the lower one is empty.  In principle, the system might
be able to relax to the ground state by emitting a phonon or a photon.
At present time, we do not have an explicit estimate for such
inelastic relaxation time, but we expect that it is quite long.  (The
other papers \cite{Riedel,Tanaka2,Tanaka-pp,Vaccarella} that
\emph{assume} thermodynamic equilibrium for each value of the phase
$\phi$ do not have an estimate of the relaxation time either.)  To
observe the predicted ac Josephson effect with the fractional
frequency $eV/\hbar$, the period of Josephson oscillations should be
set shorter than the inelastic relaxation time, but not too short, so
that the time evolution of the BdG equation can be treated
adiabatically.  Controlled nonequilibrium population of the upper
Andreev bound state was recently achieved experimentally in an
$s$-wave Josephson junction in Ref.\ \cite{Baselmans}.

Eq.\ (\ref{I_t}) can be generalized to the case where initially the
two subgap states are populated thermally at $\phi=0$, and these
occupation numbers are preserved by dynamical evolution
\begin{eqnarray}
  I_p(t) &=& {2e\over\hbar}\sum_{a}
  {\partial E_a[\phi(t)] \over \partial\phi}\, f[E_a(\phi=0)]
\label{Idyn} \\
  &=& \sin\left({eVt\over\hbar}\right)
  {\pi\Delta_0\over\sqrt{D}eR}
  \tanh\left({\Delta_0\sqrt{D}\over2T}\right).
\label{I_t-T}
\end{eqnarray}
Notice that the periodicities of the dynamical equation (\ref{I_t-T})
and the thermodynamic Eq.\ (\ref{I_p}) are different.  The latter
equation assumes that the occupation numbers of the subgap states are
in instantaneous thermal equilibrium for each $\phi$.

\subsection{Tunneling Hamiltonian approach}

In the infinite barrier limit $D\to0$, the energies $\pm E_0^{(p)}$ of
the two subgap states (\ref{E_p}) degenerate to zero, i.e.\ they
become midgap states.  The wave functions (\ref{L0s}) simplify as
follows:
\begin{eqnarray}
  \psi_{\pm0} &=& {\psi_{L0}(x)\mp\psi_{R0}(x)\over\sqrt{2}}, 
\label{pm0} \\
  \psi_{L0} &=& \sqrt{2\kappa}\,\sin(k_Fx)\,e^{\kappa x}
  \left(\begin{array}{c} 1 \\ i
  \end{array}\right)\theta(-x),
\label{L0} \\
  \psi_{R0} &=& \sqrt{2\kappa}\,\sin(k_Fx)\,e^{-\kappa x}
  \left(\begin{array}{c} e^{i\phi/2} \\ -ie^{-i\phi/2}
 \end{array}\right)\theta(x).
\label{R0} 
\end{eqnarray}
Since at $D=0$ the Josephson junction consists of two semi-infinite
uncoupled $p_x$-wave superconductors, $\psi_{L0}$ and $\psi_{R0}$ are
the wave functions of the surface midgap states \cite{Ours1} belonging
to the left and right superconductors.  Let us examine the properties
of the midgap states in more detail.

If $(u,v)$ is an eigenvector of Eq.\ (\ref{eq:BdG}) with an eigenvalue
$E_n$, then $(-v^*,u^*)$ for $s$-wave and $(v^*,u^*)$ for $p$-wave are
the eigenvectors with the energy $E_{\bar n}=-E_n$.  It follows from
these relations and Eq.\ (\ref{gamma_n}) that $\hat{\gamma}_{\bar
  n\bar\sigma\bar k_y}=C\hat{\gamma}_{n\sigma k_y}^\dag$ with $|C|=1$.
Notice that in the $s$-wave case, because $(u,v)$ and $(-v^*,u^*)$ are
orthogonal for any $u$ and $v$, the states $n$ and $\bar n$ are always
different.  However, in the $p$-wave case, the vectors $(u,v)$ and
$(v^*,u^*)$ may be proportional, in which case they describe the same
state with $E=0$.  The states (\ref{L0}) and (\ref{R0}) indeed have
this property:
\begin{equation}
  v_{L0} = iu_{L0}^*, \qquad v_{R0} = -iu_{R0}^*.
\label{uv0}
\end{equation}
Substituting Eq.\ (\ref{uv0}) into Eq.\ (\ref{gamma_n}), we find the
Bogoliubov operators of the left and right midgap states
\begin{equation}
  \hat{\gamma}^\dag_{L0\sigma k_y}
  =i\hat{\gamma}_{L0\bar\sigma\bar k_y},
  \quad
  \hat{\gamma}^\dag_{R0\sigma k_y}
  =-i\hat{\gamma}_{R0\bar\sigma\bar k_y}.
\label{conjugate}
\end{equation}
Operators (\ref{conjugate}) correspond to the Majorana fermions
discussed in Ref.\ \cite{Kitaev}.  In the presence of a midgap state,
the sum over $n$ in Eq.\ (\ref{canon}) should be understood as
$\sum_{n>0}+(1/2)\sum_{n=0}$, where we identify the second term as the
projection ${\cal P}\hat c$ of the electron operator onto the midgap
state.  Using Eqs.\ (\ref{uv0}), (\ref{conjugate}), and (\ref{canon}),
we find
\begin{equation}
  {\cal P}\hat c_{\sigma k_y}(x) 
  = u_0(x)\hat\gamma_{0\sigma k_y}
  = v_0^*(x)\hat\gamma_{0\bar\sigma\bar k_y}^\dag.
\label{c_0}
\end{equation}

Let us consider two semi-infinite $p_x$-wave superconductors on a 1D
lattice with the spacing $l$, one occupying $x\le\bar l=-l$ and
another $x\ge l$.  They are coupled by the tunneling matrix element
$\tau$ between the sites $\bar l$ and $l$:
\begin{equation}
  \hat H_\tau = \tau \sum_{\sigma k_y} 
  [ \hat{c}^\dag_{L\sigma k_y}(\bar l)\,\hat{c}_{R\sigma k_y}(l) + 
  \hat{c}^\dag_{R\sigma k_y}(l)\,\hat{c}_{L\sigma k_y}(\bar l) ]. 
\label{HT}
\end{equation}
In the absence of coupling ($\tau=0$), the subgap wave functions of
each superconductor are given by Eqs.\ (\ref{L0}) and (\ref{R0}).
Using Eqs.\ (\ref{c_0}), (\ref{uv0}), (\ref{L0}), and (\ref{R0}), the
tunneling Hamiltonian projected onto the basis of midgap states is
\begin{eqnarray}
  && {\cal P}\hat H_\tau = \tau \,
  [u_{L0}^*(\bar l) u_{R0}(l) + {\rm c.c.}] \,
  (\hat\gamma_{L0\uparrow}^\dag \hat\gamma_{R0\uparrow}
  + {\rm H.c.})
\nonumber \\
  && = \Delta_0\sqrt{D}\, \cos(\phi/2)\,
  (\hat\gamma_{L0\uparrow}^\dag \hat\gamma_{R0\uparrow}
  + \hat\gamma_{R0\uparrow}^\dag \hat\gamma_{L0\uparrow}),
\label{PHT}
\end{eqnarray}
where $\sqrt{D}=4\tau\sin^2k_Fl/\hbar v_F$ is the transmission
amplitude, and we omitted summation over the diagonal index $k_y$.
Notice that Eq.\ (\ref{PHT}) is $4\pi$-periodic in $\phi$
\cite{Kitaev}.

Hamiltonian (\ref{PHT}) operates between the two degenerate states of
the system related by annihilation of the Bogoliubov quasiparticle in
the right midgap state and its creation in the left midgap state.  In
this basis, Hamiltonian (\ref{PHT}) can be written as a $2\times2$
matrix
\begin{equation}
  {\cal P}\hat H_\tau = \Delta_0\sqrt{D}\cos(\phi/2)
  \left( \begin{array}{cc} 0 & 1 \\ 1 & 0 \end{array} \right). 
\label{two-level}
\end{equation}
The eigenvectors of Hamiltonian (\ref{two-level}) are $(1,\mp1)$,
i.e.\ the antisymmetric and symmetric combinations of the right and
left midgap states given in Eq.\ (\ref{pm0}).  Their eigenenergies are
$E_\pm(\phi)=\mp\Delta_0\sqrt{D}\cos(\phi/2)$, in agreement with Eq.\ 
(\ref{E_p}).  The tunneling current operator is obtained by
differentiating Eqs.\ (\ref{PHT}) or (\ref{two-level}) with respect to
$\phi$.  Because $\phi$ appears only in the prefactor, the operator
structures of the current operator and the Hamiltonian are the same,
so they are diagonal in the same basis.  Thus, the energy eigenstates
are simultaneously the eigenstates of the current operator with the
eigenvalues
\begin{equation}
  I_\pm=\pm{\sqrt{D}e\Delta_0\over\hbar}
  \sin\left({\phi\over2}\right),
\label{I_pm}
\end{equation}
in agreement with Eq.\ (\ref{I_t}).  The same basis $(1,\mp1)$
diagonalizes Hamiltonian (\ref{two-level}) even when a voltage $V$ is
applied and the phase $\phi$ is time-dependent.  Then the initially
populated eigenstate with the lower energy produces the current
$I_p=\sqrt{D}(e\Delta_0/\hbar)\sin(eVt/\hbar)$ with the fractional
Josephson frequency $eV/\hbar$, in agreement with Eq.\ (\ref{I_t}).

\subsection{Josephson current carried by single electrons, 
  rather than Cooper pairs}
\label{sec:single}

In the tunneling limit, the transmission coefficient $D$ is
proportional to the square of the electron tunneling amplitude $\tau$:
$D\propto\tau^2$.  Eqs.\ (\ref{I_t}) and (\ref{I_pm}) show that the
Josephson current in the $p_x$-$p_x$ junction is proportional to the
first power of the electron tunneling amplitude $\tau$.  This is in
contrast to the $s$-$s$ junction, where the Josephson current
(\ref{I_s}) is proportional to $\tau^2$.  This difference results in
the big ratio $I_p/I_s=2/\sqrt{D}$ between the critical currents at
$T=0$ in the $p_x$- and $s$-wave cases, as shown in Fig.\ \ref{fig:Ic}
and discussed in Sec.\ \ref{sec:dc}.  The reason for the different
powers of $\tau$ is the following.  In the $p_x$-wave case, the
transfer of just one electron between the degenerate left and right
midgap states is a real (nonvirtual) process.  Thus, the eigenenergies
are determined from the secular equation (\ref{two-level}) already in
the first order of $\tau$.  In the $s$-wave case, there are no midgap
states, so the transferred electron is taken from below the gap and
placed above the gap, at the energy cost $2\Delta_0$.  Thus, the
transfer of a single electron is a virtual (not real) process.  It
must be followed by the transfer of another electron, so that the pair
of electrons is absorbed into the condensate.  This gives the current
proportional to $\tau^2$.

This picture implies that the Josephson supercurrent across the
interface is carried by single electrons in the $p_x$-$p_x$ junction
and by Cooper pairs in the $s$-$s$ junction.  Because the
single-electron charge $e$ is a half of the Cooper-pair charge $2e$,
the frequency of the ac Josephson effect in the $p_x$-$p_x$ junction
is $eV/\hbar$, a half of the conventional Josephson frequency
$2eV/\hbar$ for the $s$-$s$ junction.  These conclusions also apply to
a junction between two cuprate $d$-wave superconductors in such
orientation that both sides of the junction have surface midgap
states, e.g. to the $45^\circ/45^\circ$ junction (see Sec.\ 
\ref{sec:d-wave}).

In both the $p_x$-$p_x$ and $s$-$s$ junctions, electrons transferred
across the interface are taken away into the bulk by the supercurrent
of Cooper pairs.  In the case of the $p_x$-$p_x$ junction, a single
transferred electron occupies a midgap state until another electron
gets transferred.  Then the pair of electrons becomes absorbed into
the bulk condensate, the midgap state returns to the original
configuration, and the cycle repeats.  In the case of the $s$-$s$
junction, two electrons are simultaneously transferred across the
interface and become absorbed into the condensate.  Clearly, electric
charge is transferred across the interface by single electrons at the
rate proportional to $\tau$ in the first case and by Cooper pairs at
the rate proportional to $\tau^2$ in the second case, but the bulk
supercurrent is carried by the Cooper pairs in both cases.

\subsection{Josephson effect between triplet superconductors with 
  nonparallel $\bm{n}$-vectors}
\label{sec:nonparallel}

In this section, we consider the Josephson effect between two
$p_x$-wave superconductors with nonparallel spin-polarization vectors
$\bm{n}$ forming an angle $\theta$.  This problem was studied in Ref.\ 
\cite{Vaccarella} using a tunneling Hamiltonian approach.  Here we
analyze the problem using the BdG formulation.  There are experimental
indications that the spin-polarization vector $\bm{n}$ is parallel to
the crystal axis $\bm{c}$ in the $\rm(TMTSF)_2X$ compounds
\cite{Chaikin,Lebed}.  Then the considered junction can be realized in
the geometry shown in Fig.\ \ref{fig:JJ}(a) where the $\bm{c}$ axes of
the two superconductors are rotated relative to each other by the
angle $\theta$ around the common $\bm{a}$ axis along the chains.

Let us select the spin quantization axis $z$ perpendicular to both
vectors $\bm{n}$, and the $x$ axis in the spin space parallel to the
vector $\bm{n}$ of the left superconductor.  Then the vector $\bm{n}$
of the right superconductor lies in the $(x,y)$ plane at the angle
$\theta$ to the $x$ axes: $\bm{n}=(\cos\theta,\sin\theta,0)$.  In this
representation, the superconducting pairing takes place between
electrons with parallel spins:
\begin{eqnarray}
  \langle\hat{c}_\sigma(\bm{k})\hat{c}_{\sigma'}(-\bm{k})\rangle
  & \propto & 
  i\hat\sigma^{(y)}\,(\hat\sigma^{(x)} n_x + \hat\sigma^{(y)} n_y)
  \,\Delta(\bm{k}) 
\nonumber \\
  & = & \left(
  \begin{array}{cc} e^{i\theta} & 0 \\ 
  0 & -e^{-i\theta} \end{array}
  \right)  \Delta(\bm{k}).
\label{eq:theta}  
\end{eqnarray}
Then, the Josephson effect can be considered separately for the spin
up and down sectors having the phase differences $\phi\pm\theta$,
correspondingly.  Using Eq.\ (\ref{E_p}) for the $p_x$-$p_x$ junction,
we find the energies of the Andreev bound states for each spin sector
\begin{eqnarray}
  E_{a,\uparrow} &=& -a \Delta_0 \sqrt{D} 
  \cos\left({\phi+\theta \over 2}\right), 
\label{E_up} \\
  E_{a,\downarrow} &=& -a \Delta_0  \sqrt{D} 
  \cos\left({\phi-\theta \over 2}\right).
\label{E_down}
\end{eqnarray}

The total Josephson current is obtained by adding the currents carried
by the two spin sectors \cite{double}.  For simplicity, below we
consider only the case of zero temperature.  In the dynamical limit,
assuming that the states (\ref{E_up}) and (\ref{E_down}) with $a=+$
are occupied initially and the occupation numbers are preserved
dynamically and using Eq.\ (\ref{I_t}), we find a $4\pi$-periodic
current:
\begin{eqnarray}
  I(t) &=& {\sqrt{D}e\Delta_0 \over 2\hbar}
  \left[ \sin\left({\phi+\theta \over 2}\right) 
  + \sin\left({\phi-\theta \over 2}\right) \right]
\nonumber \\
  &=& {\sqrt{D}e\Delta_0 \over \hbar}\,
  \sin\left({\phi(t) \over 2}\right)\,
  \cos\left({\theta \over 2}\right).
\label{I-theta-t}
\end{eqnarray}
In the static thermodynamic limit, using Eq.\ (\ref{I_p}) at $T=0$, we
find the dc Josephson current:
\begin{eqnarray}
  I &=& {\sqrt{D}e\Delta_0\over 2\hbar}
  \left\{ \sin\left({\phi+\theta \over 2}\right)\,
  {\rm sgn}\left[ \cos\left({\phi+\theta \over 2}\right)
  \right] \right.
\nonumber \\
  && + \left. \sin\left({\phi-\theta \over 2}\right)\,
  {\rm sgn}\left[ \cos\left({\phi-\theta \over 2}\right)
  \right] \right\}.
\label{I-theta}
\end{eqnarray}

For completeness, let us also consider the Josephson effect between
two $p_y$-wave or two $p_z$-wave superconductors, where the $y$ and
$z$ axes are parallel to the junction plane.  In these junctions,
midgap states are absent in the $D\to0$ limit, thus the current-phase
relation is conventional $I=I_c\sin\phi$.  For nonparallel vectors
$\bm{n}$, the total Josephson current is the sum of the spin up and
down sectors:
\begin{eqnarray}
  I &=& {I_c \over 2}\, [\sin(\phi+\theta)+\sin(\phi-\theta)] 
\nonumber \\
  &=& I_c\cos\theta\sin\phi = I_c(\bm{n}_L\cdot\bm{n}_R)\sin\phi.
\label{cos-theta}
\end{eqnarray}
Eq.\ (\ref{cos-theta}) is consistent with Ref.\ \cite{Vaccarella}.  In
the case where the two vectors $\bm{n}$ are perpendicular
($\theta=\pi/2$), the Josephson current (\ref{cos-theta}) for the
superconductors without midgap states vanishes, but, according to
Eqs.\ (\ref{I-theta-t}) and (\ref{I-theta}), it is not zero if the
midgap states are present.

\subsection{\boldmath $s$-$p_x$ junction between singlet and triplet 
  superconductors}
\label{sec:s-p}

In this section, we consider a junction between a singlet $s$-wave and
a triplet $p_x$-wave superconductors.  The junction geometry is the
same as in Fig.\ \ref{fig:JJ}(a), where one of the superconductors is
taken to be a conventional $s$-wave superconductor and another one a
Q1D triplet $p_x$-wave superconductor, such as $\rm(TMTSF)_2X$.

\begin{figure}[t] 
\includegraphics[width=0.6\linewidth,angle=0]{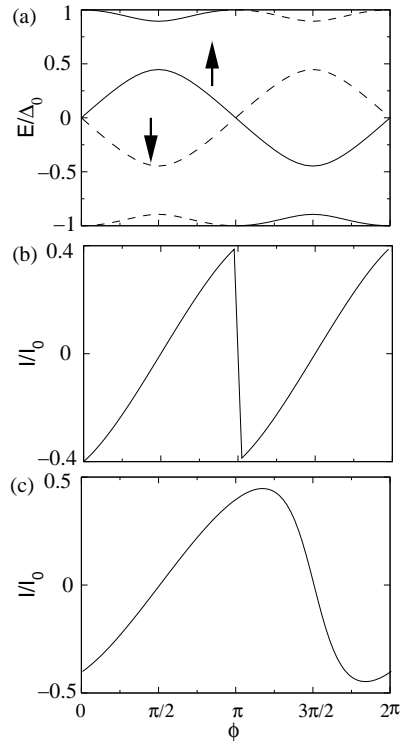} 
\caption{ The subgap energy levels and the Josephson currents in the 
  $s$-$p_x$ junction.  Here $D=0.8$ and $I_0=e\Delta_0 /2\hbar$.  (a)
  The energies (\ref{sp-E}) of the Andreev bound states.  The solid
  and dashed lines correspond to $\sigma=\uparrow$ and
  $\sigma=\downarrow$.  (b) The Josephson current in the static
  thermodynamic limit, where the states with $E<0$ are occupied.  (c)
  The Josephson current in the dynamical limit, where the central
  branch with $\sigma=\downarrow$ and the lower branches touching
  $-\Delta_0$ are occupied.}
\label{fig:spj}
\end{figure}

We choose the spin quantization axis $z$ along the polarization vector
$\bm{n}$ of the triplet superconductor, so the spin projection
$\sigma$ on the $z$ axis is a good quantum number.  In both triplet
and singlet superconductors, the Cooper pairing takes place between
electrons with opposite spins.  However, the pairing potential has the
same sign for $\sigma$ and $\bar\sigma$ in the triplet superconductor
and the opposite signs in the singlet superconductor.  Thus, the phase
difference across the Josephson junction is $\phi$ for quasiparticles
with $\sigma=\uparrow$ and $\phi+\pi$ for $\sigma=\downarrow$.  The
energies of the Andreev bound states can be found for each $\sigma$
from Eq.\ (\ref{self}) together with Eq.\ (\ref{kappa-eta}), where we
should use the upper line of Eq.\ (\ref{Delta_bsa}) for the left
superconductor and the lower line for the right superconductor.  To
simplify calculations, we consider the case where the magnitudes of
the gaps are equal for the $s$- and $p_x$-wave superconductors:
$|\Delta_L|=|\Delta_R|=\Delta_0$.  The energies of the Andreev bound
states are
\begin{equation}
  E_{a,\sigma} =  \sigma\,{\rm sgn}(\sin\phi) 
  \Delta_0 \sqrt{1+a\sqrt{1-D^2\sin^2\phi} \over 2}.
\label{sp-E}
\end{equation}
For each value of the spin index $\sigma=\pm$, Eq.\ (\ref{sp-E}) gives
two Andreev states labeled by the index $a=\pm$.  In the tunneling
limit $D\ll1$, we have
\begin{eqnarray}
  E_{+,\sigma} &\approx& \sigma\,{\rm sgn}(\sin\phi)\,\Delta_0
  \left(1-\frac18\, D^2\sin^2\phi\right),
\label{eq:E+} \\
  E_{-,\sigma} &\approx& \frac12\, \sigma\Delta_0 D\sin\phi.
\label{eq:E-}
\end{eqnarray}
The energies (\ref{sp-E}) are plotted in Fig.\ \ref{fig:spj}(a) vs.\ 
$\phi$ by the solid lines for $\sigma=\uparrow$ and by the dashed
lines for $\sigma=\downarrow$.  We observe that the branches
(\ref{eq:E+}) with $a=+$ touch the gap boundaries $\pm\Delta_0$ at
$\phi=0$ and $\pi$, whereas the branches (\ref{eq:E-}) with $a=-$ stay
in the center of the gap.

In the limit $D\ll1$, the central branches with $a=-$ dominate the
energy dependence on $\phi$, and energy minima are achieved at
$\phi=\pi/2$ or $3\pi/2$.  Notice that if the system selects the
energy minimum at $\phi=\pi/2$, then the spin down states, shown by
the dashed lines in Fig.\ \ref{fig:spj}(a), are populated, and the
spin up states are empty, so the junction accumulates the spin
$-\hbar/2$ per conducting channel \cite{double}.  If the system
selects the energy minimum at $\phi=3\pi/2$, then the junction
accumulates the spin $\hbar/2$ per conducting channel.

In the limit $D\ll1$, we can neglect the energies (\ref{eq:E+}) and
obtain the Josephson current by differentiating the energies
(\ref{eq:E-}) with respect to $\phi$ using Eq.\ (\ref{eq:I_a})
\cite{double}.  In the dynamical limit, the occupation numbers of the
Andreev states (\ref{eq:E-}) are preserved, and the Josephson current
has the $2\pi$-periodicity, as shown in Fig.\ \ref{fig:spj}(c):
\begin{equation}
  I(t) \approx -{e\over 2\hbar}\Delta_0 D\cos\phi(t).
\end{equation}
In the static thermodynamic limit, the system occupies the branch of
the minimal energy for each $\phi$, and the Josephson current is
$\pi$-periodic, as shown in Fig.\ \ref{fig:spj}(b):
\begin{equation}
  I(\phi) \approx -{e\over 2\hbar}
  \Delta_0 D\,{\rm sgn}(\sin\phi)\,\cos\phi.
\label{I_sp}
\end{equation}
The thermodynamic assumption implies that the spin accumulation at the
$s$-$p_x$ junction changes sign when the phase $\phi$ crosses $\pi$.

\begin{figure}[t] 
\includegraphics[width=0.6\linewidth,angle=0]{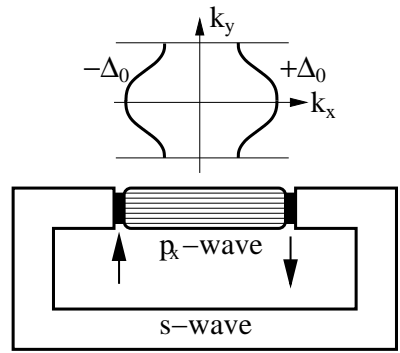} 
\caption{A Q1D $p_x$-wave superconductor closed in a loop by an 
  $s$-wave superconductor.  No current is circulating in the loop in
  equilibrium.  However, there is accumulation of spins up in one
  $s$-$p_x$ junction and spins down in another junction.  The sketch
  at the top illustrates the Fermi surface of a Q1D metal with the
  opposite signs of the superconducting $p_x$-wave pairing potential
  on the two sheets of the Fermi surface.}
\label{fig:sp-loop}
\end{figure}

Now let us consider the circuit shown in Fig.\ \ref{fig:sp-loop},
where a $p_x$-wave superconductor has Josephson junctions at both ends
with an $s$-wave superconductor closed in a loop.  Because the sign of
the $p_x$-wave pairing potential is opposite for the $+k_F$ and $-k_F$
sheets of the Fermi surface, the two junctions have the relative phase
shift $\pi$.  Naively, one might expect a spontaneous current in this
circuit by analogy with the corner SQUID in the cuprates
\cite{vanHarlingen}.  However, the system shown in Fig.\ 
\ref{fig:sp-loop} can accommodate the phase shift $\pi$ by selecting
the energy minimum at $\phi=\pi/2$ for one junction and the energy
minimum at $\phi=3\pi/2$ for another junction.  Then, no current
circulates in the loop.  However, one junction accumulates spins up
and another junction spins down, which might be possible to detect
experimentally.

The results of the this section clearly show that a Josephson current
is possible between singlet and triplet superconductors, in agreement
with the earlier findings by Yip \cite{Yip}.  Recently, the Josephson
current was calculated for the $s$-$p_x$ junction in Ref.\ 
\cite{Asano}, but spin accumulation at the junction was not recognized
in this paper.  The $s$-$p_x$ junction considered in this section is
mathematically equivalent to the $0^\circ/45^\circ$ $d$-$d$ junction
and the $45^\circ$ junction between an $s$-wave and a $d$-wave
superconductors (see Sec.\ \ref{sec:d-wave}).  Eq.\ (\ref{sp-E}) was
obtained for that case in Refs.\ \cite{Riedel,Barash,Tanaka2}.
However, there is no spin accumulation in junctions between singlet
$s$- and $d$-wave superconductors, unlike in the $s$-$p_x$ junction.

\section{Junctions between quasi-two-dimensional superconductors}

In this section, we study junctions between quasi-two-dimensional
(Q2D) superconductors such as nonchiral $d$-wave cuprates and chiral
$p_x\pm ip_y$-wave ruthenates.  For simplicity, we use an isotropic
electron energy dispersion law
$\varepsilon=\hbar^2(k_x^2+k_y^2)/2m-\mu$ in the $(x,y)$ plane.  As
before, we select the coordinate $x$ perpendicular to the junction
line and assume that the electron momentum component $k_y$ parallel to
the junction line is a conserved good quantum number.  Then, the 2D
problem separates into a set of 1D solutions (\ref{L0s}) in the $x$
direction labeled by the index $k_y$.  The Fermi momentum $k_F$ and
velocity $v_F$ are replaced by their $x$-components
$k_{Fx}=\sqrt{k_{F}^2-k_y^2}$ and $v_{Fx}=\hbar k_{Fx}/m$.  The
transmission coefficient of the barrier (\ref{ZDG}) becomes
$k_y$-dependent
\begin{equation}
  Z(k_y)=Z_0{k_F \over \sqrt{k_F^2-k_y^2}}, \quad 
  D(k_y)={4 \over Z^2(k_y)+4},
\label{ZDG-2D}
\end{equation}
where $Z_0=\sqrt{2mU_0}/\hbar k_F$.  The total Josephson current is
given by a sum over all occupied subgap states labeled by $k_y$.

\subsection{Josephson junctions between $d$-wave superconductors}
\label{sec:d-wave}

\begin{figure}[t] 
\includegraphics[width=\linewidth,angle=0]{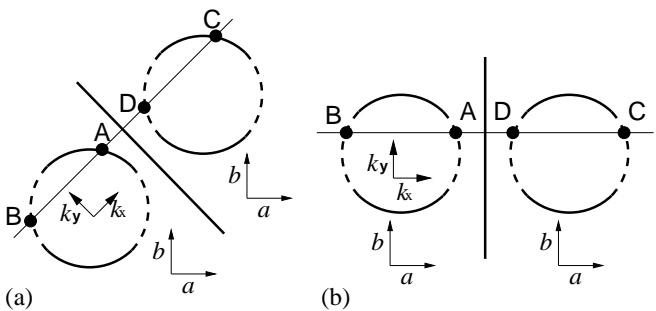} 
\caption{Schematic drawing of the $45^\circ/45^\circ$ junction 
  (panel a) and $0^\circ/0^\circ$ junction (panel b) between two
  $d$-wave superconductors.  The thick line represents the junctions
  line.  The circles illustrate the Fermi surfaces, where positive and
  negative pairing potentials $\Delta$ are shown by the solid and
  dotted lines.  The points A, B, C, and D in the momentum space are
  connected by transmission and reflection from the barrier.}
\label{fig:d-wave}
\end{figure}

For the cuprates, let us consider a junction parallel to the
$[1,\bar1]$ crystal direction in the $(\bm{a},\bm{b})$ plane and
select the $x$ axis along the diagonal $[1,1]$, as shown in Fig.\ 
\ref{fig:d-wave}(a).  In these coordinates, the $d$-wave pairing
potential is
\begin{equation}
  \hat\Delta_{\sigma k_y}(x,\hat k_x) =
  \sigma2\Delta_\beta\,k_y \hat k_x/k_F^2,
\label{hat-Delta-d}
\end{equation}
where the same notation as in Eq.\ (\ref{hat-Delta}) is used.  Direct
comparison of Eqs.\ (\ref{hat-Delta-d}) and (\ref{hat-Delta})
demonstrates that the $d$-wave superconductor with the $45^\circ$
junction maps to the $p_x$-wave superconductor by the substitution
$\Delta_0\to\sigma2\Delta_0k_y /k_F$.  Thus, the results obtained in
Sec.\ \ref{sec:Q1D} for the $p_x$-$p_x$ junction apply to the
$45^\circ/45^\circ$ junction between two $d$-wave superconductors with
the appropriate integration over $k_y$.  The energies of the subgap
Andreev states are given by Eq.\ (\ref{E_p}) with the $k_y$-dependent
parameters $\Delta_0$ and $D$, and the energies and the wave functions
are $4\pi$-periodic functions of $\phi$.  Thus, the ac Josephson
current in the dynamical limit is $4\pi$-periodic and has the
fractional frequency $eV/\hbar$, as in Eqs.\ (\ref{I_t}),
(\ref{I_t-T}), and (\ref{I_pm}).  The energies (\ref{E_p}) of the
subgap states \cite{Tanaka1,Riedel} and the dc Josephson current
(\ref{I_p}) in the thermodynamic limit \cite{Riedel,Tanaka2} were
calculated for the $45^\circ/45^\circ$ $d$-$d$ junction before.
However, these papers did not recognize the fractional,
$4\pi$-periodic character of the Josephson effect in the dynamical
limit.

On the other hand, if the junction is parallel to the $[0,1]$ crystal
direction, as shown in Fig.\ \ref{fig:d-wave}(b), then
$\hat\Delta_{\sigma k_y}(x,\hat k_x) = \sigma\Delta_\beta\, (\hat
k_x^2-k_y^2)/k_F^2$.  This pairing potential is an even function of
$\hat k_x$, thus it is analogous to the $s$-wave pairing potential in
Eq.\ (\ref{hat-Delta}).  Then, the $0^\circ/0^\circ$ junction between
two $d$-wave superconductors is analogous to the $s$-$s$ junction.  It
should exhibit the conventional $2\pi$-periodic Josephson effect with
the frequency $2eV/\hbar$.

For a generic orientation of the junction line, the $d$-wave pairing
potential is $p_x$-like for some momenta $k_y$ and $s$-like for other
$k_y$.  Thus, the total Josephson current is a sum of the
unconventional and conventional terms:
\begin{equation}
  I=C_0\sin(\phi/2) + C_1\sin(\phi) + \ldots~,
\label{I-phi/2}  
\end{equation}
where $C_0$ and $C_1$ are some coefficients.  We expect that both
terms in Eq.\ (\ref{I-phi/2}) are present for any real junction
between $d$-wave superconductors because of imperfections in junction
orientation.  However, the ratio $C_0/C_1$ should be maximal for the
junction shown in Fig.\ \ref{fig:d-wave}(a) and minimal for the
junction shown in Fig.\ \ref{fig:d-wave}(b).  In general, whenever the
superconductors on both sides of the junction have surface midgap
states, we expect to observe the $4\pi$-periodic fractional ac
Josephson effect.  In principle, the effect may be spoiled by the
gapless quasiparticles that exist near the gap nodes in a $d$-wave
superconductor.  However, they would affect only a small portion of
the Fermi surface near the nodes, and the $4\pi$-periodic Josephson
effect should survive on the other parts of the Fermi surface, where
the gap is big.

The $45^\circ/45^\circ$ junction shown in Fig.\ \ref{fig:d-wave}(a)
should not be confused with the $0^\circ/45^\circ$ $d$-$d$ junction
\cite{Wendin-ac} or the $45^\circ$ $s$-$d$ junction
\cite{Tanaka-sd,Zagoskin-sd}, much discussed in literature.  None of
the papers \cite{Tanaka-sd,Zagoskin-sd,Wendin-ac} treated the problem
correctly, because they did not take into account the Andreev bound
states in the junction properly.  The correct energy spectrum of the
Andreev bound states was obtained in Refs.\ 
\cite{Tanaka1,Riedel,Barash}.  In the $0^\circ/45^\circ$ $d$-$d$ and
$45^\circ$ $s$-$d$ junctions, only one superconductor has midgap
states, thus these junctions are mathematically analogous to the
$s$-$p_x$ junction considered in Sec.\ \ref{sec:s-p}.  The spectrum of
the Andreev bound states is given by Eq.\ (\ref{sp-E}) without the
factor $\sigma$, because both superconductors are singlet.  The energy
levels are plotted vs.\ $\phi$ in Fig.\ \ref{fig:spj}(a), where the
solid and dashed lines represent not spin, but positive and negative
momenta $k_y$.  The junction has two energy minima at $\phi=\pi/2$ or
$3\pi/2$, where the states with only negative or positive momenta
$k_y$ are filled, thus there are persistent currents along the
junction line \cite{Huck,Amin}.  (On the other hand, there is no spin
accumulation, unlike in the $s$-$p_x$ junction discussed in Sec.\ 
\ref{sec:s-p}.)  In the thermodynamic limit, the current-phase
relation shown in Fig.\ \ref{fig:spj}(b) is $\pi$-periodic; however,
it requires reversing the currents along the junction line when $\phi$
passes through 0 or $\pi$.  In the dynamical limit, the current-phase
relation shown in Fig.\ \ref{fig:spj}(c) is $2\pi$-periodic.  The
first two harmonics $I=C_1\sin(\phi)+C_2\sin(2\phi)$ have been
recently observed experimentally in the $0^\circ/45^\circ$ $d$-$d$
junction \cite{Lindstrom}.

\subsection{Josephson junctions with chiral superconductors}

In this section, we study junctions between the chiral $p_x\pm
ip_y$-wave superconductors $\rm Sr_2RuO_4$, where the pairing
potential is assumed to be $\Delta({\bf k})=\Delta_0(k_x \pm
ik_y)/k_F$ \cite{Maeno}, and the two signs correspond to opposite
chiralities.  We assume a uniform orientation of the spin-polarization
vector $\bm{n}$ across the junction.  This problem was investigated in
Ref.\ \cite{Barash-chiral} using the Eilenberger equation for Green's
functions.  It was found that the chiral subgap states at the
junctions enhance the low-temperature critical Josephson current in
symmetric junctions.  Here we use the BdG equation to obtain the
spectrum of the Andreev bound states.  As before, we assume that the
momentum component $k_y$ parallel to the junction is conserved.  Thus,
the problem separates into a set of 1D solutions in the $x$ direction
perpendicular to the junction plane, and we can use the method of
Sec.\ \ref{sec:bound}.

\begin{widetext}
\begin{figure}[b]
\includegraphics[width=2\linewidth,angle=0]{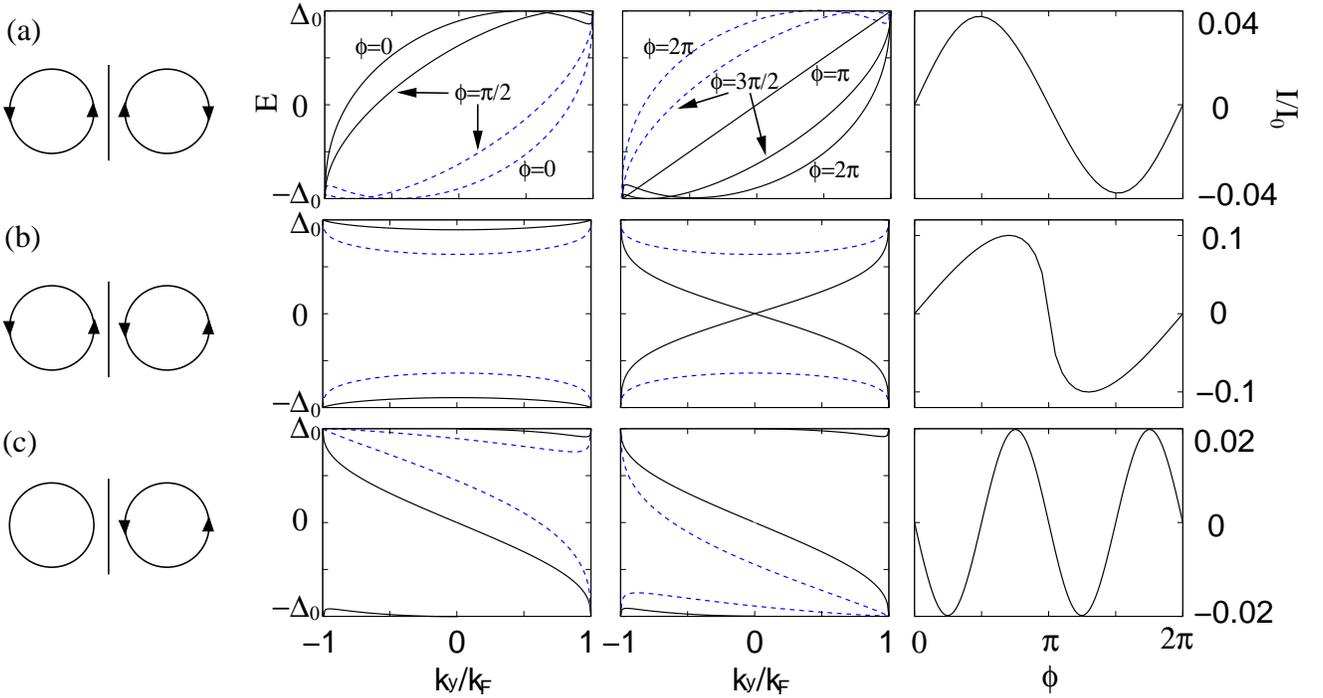}
\caption{
  The subgap energy spectra and the Josephson currents in the
  junctions with chiral $p_x\pm ip_y$-wave superconductors, calculated
  for $Z_0=1$.  (a) Junction between two $p$-wave superconductors with
  opposite chiralities.  (b) Junction between two $p$-wave
  superconductors with the same chirality.  (c) Junction between
  $s$-wave and chiral $p$-wave superconductors.  In the second and
  third columns of row (a), the solid and dashed lines show the
  $a=\pm$ branches of Eq.\ (\ref{opposite-ch}) for different values of
  $\phi$.  In rows (b) and (c), the energies (\ref{same-ch}) and
  (\ref{E-sp}) are shown by the solid and dashed lines for $\phi=0$
  and $\pi/2$ in the second column and for $\phi=\pi$ and $3\pi/2$ in
  the third column.  The energies of quasiparticles with
  $\sigma=\downarrow$ are the same as with $\sigma=\uparrow$ for rows
  (a) and (b), and they can be obtained by the shift $\phi\to\phi+\pi$
  for row (c).  The fourth column shows the Josephson current in the
  static thermodynamic limit, normalized by
  $I_0=e\Delta_0L_yk_F/2\pi\hbar$, where $L_y$ is the length of the
  junction.}
\label{fig:chiral}
\end{figure}
\end{widetext}

First we consider a junction between two superconductors with opposite
chiralities, as illustrated in the first column of Fig.\ 
\ref{fig:chiral}(a).  In this case, $\Delta_L=\Delta_0(k_x+ik_y)/k_F$
and $\Delta_R=e^{i\phi}\Delta_0(k_x-ik_y)/k_F$.  When the barrier is
not transparent ($D=0$), each superconductor has chiral Andreev
surface states with the same energy dispersion
$E(k_y)=k_y\Delta_0/k_F$ \cite{Honerkamp98}.  The electron tunneling
amplitude $\tau\propto\sqrt{D}$ produces a matrix element mixing the
two states in the first-order degenerate perturbation theory.  Thus,
the two energy spectra repel with the splitting proportional to
$\sqrt{D}$.  From Eq.\ (\ref{self}), we find the following subgap
energies:
\begin{equation}
  E = {\Delta_0 \over k_F} \left( 
  k_y \sqrt{1-D\cos^2{\phi\over2}}  
  \pm \sqrt{k_F^2-k_y^2} \sqrt{D} \cos{\phi\over2}
  \right).
\label{opposite-ch}
\end{equation}
The energy levels splitting oscillates with the period $4\pi$ as a
function of $\phi$: $\delta
E=(\Delta_0/k_F)\sqrt{k_F^2-k_y^2}\sqrt{D}\cos\phi/2$.  The splitting
depends on $k_y$ through the square-root prefactor and through the
dependence of $D$ on $k_y$ in Eq.\ (\ref{ZDG-2D}), and vanishes at
$k_y =\pm k_F$.  The energy dispersion (\ref{opposite-ch}) is plotted
vs.\ $k_y$ in the second and third columns of Fig.\ 
\ref{fig:chiral}(a) for several values of $\phi$.  The spectrum of
excitations is gapless because of the chiral dispersion in $k_y$.
Thus, it is reasonable to assume that the occupation numbers of the
subgap states are in instantaneous thermodynamic equilibrium for any
phase $\phi$.  Then, the Josephson current is a $2\pi$-periodic
function of $\phi$, as illustrated at zero temperature in the fourth
column of Fig.\ \ref{fig:chiral}(a), even though the energy levels
(\ref{opposite-ch}) are $4\pi$-periodic functions of $\phi$.

Now let us consider the case of two superconductors with opposite
chiralities, as illustrated in the first column of Fig.\ 
\ref{fig:chiral}(b).  When the two superconductors are disconnected
($D=0$), their chiral Andreev surface states have opposite dispersions
$E=\pm k_y\Delta_0/k_F$, thus they are nondegenerate.  The electron
tunneling amplitude $\tau\propto\sqrt{D}$ repels the energy levels
around the intersection point $k_y=0$.  From Eq.\ (\ref{self}), we
find the following subgap energies:
\begin{equation}
  E = \pm {\Delta_0 \over k_F} 
  \sqrt{(1-D)k_y^2 + D k_F^2\cos^2{\phi\over2}}.
\label{same-ch}
\end{equation}
The energy dispersion (\ref{same-ch}) is plotted vs.\ $k_y$ in the
second and third columns of Fig.\ \ref{fig:chiral}(b) for several
values of $\phi$.  The energy splitting around $k_y=0$ is a
$2\pi$-periodic function of $\phi$ and vanishes at $\phi=\pi$.  The
Josephson current is a $2\pi$-periodic function of $\phi$, as
illustrated at zero temperature in the fourth column of Fig.\ 
\ref{fig:chiral}(b).

Now let us consider a junction between an $s$-wave and a
$p_x+ip_y$-wave superconductors shown in the first column of Fig.\ 
\ref{fig:chiral}(c).  The Josephson current was calculated in this
case in Ref.\ \cite{Asano} using the method of Green's functions.
However, the energies of the Andreev bound states were not written
explicitly.  The subgap states in this junction are obtained by
solving Eq.\ (\ref{self}) in the manner similar to the 1D $s$-$p_x$
junction.  For simplicity, we assume that the magnitudes of the
pairing potentials in both superconductors are the same:
$|\Delta_L|=|\Delta_R|=\Delta_0$.  The square of the subgap energies
is given by the following expression
\begin{eqnarray}
  E_{a,\sigma}^2 &=& {\Delta_0^2 \over 2} 
  \bigg( 1 + R\tilde k_y^2 - \sigma D \tilde k_y\sin\phi
\label{E-sp} \\
  && \left. {} + a\sqrt{1 - \tilde k_y^2} 
  \sqrt{1 - (R\tilde k_y - \sigma D\sin\phi)^2}
  \right),
\nonumber
\end{eqnarray}
where $R=1-D$ is the reflection coefficient, and $\tilde k_y=k_y/k_F$.
The signs of the energies are
\begin{eqnarray}
  {\rm sgn}E_{a,\sigma} &=& {\rm sgn}
  \bigg( -R\tilde k_y + \sigma D \sin\phi 
\label{signs} \\
  && \left. {} + a \tilde k_y
  \sqrt{1 - (R\tilde k_y - \sigma D\sin\phi)^2 \over 1-\tilde k_y^2}
  \right).
\nonumber
\end{eqnarray}
For a given $\sigma$, there are two branches of energies labeled by
the index $a=\pm$.  The energy dispersions $E_{\pm,\uparrow}(k_y)$ are
shown in the second and third columns of Fig.\ \ref{fig:chiral}(c) for
several phases $\phi$.  In the limit of impenetrable barrier $D\to0$,
the energy branch with $a=+$ approaches to the gap edges
$|E_+|\to\Delta_0$, whereas the branch with $a=-$ approaches to the
energy dispersion $E_-\to- k_y\Delta_0/k_F$ of the chiral surface
states in the $p_x+ip_y$-wave superconductor \cite{Honerkamp98}.

The energy $E_{a,\sigma}(k_y)$ for a given $\sigma$ is a
$2\pi$-periodic function of $\phi$. The energy $E_{a,\bar\sigma}(k_y)$
is obtained from $E_{a,\sigma}(k_y)$ by the shift $\phi\to\phi+\pi$,
as discussed in Sec.\ \ref{sec:s-p}.  Thus, the Josephson current
(\ref{thermal}) in the static thermodynamic limit, obtained by
summation over $\sigma$ and $k_y$, is a $\pi$-periodic function of
$\phi$, as shown in the fourth column of Fig.\ \ref{fig:chiral}(c), in
agreement with Ref.\ \cite{Asano}.  Similarly to the $s$-$p_x$
junction considered in Sec.\ \ref{sec:s-p}, the $s$-($p_x+ip_y$)
junction has two equal energy minima \cite{Asano} at $\phi=\pi/2$ and
$3\pi/2$ accompanied by accumulation of the down spin for $\phi=\pi/2$
and the up spin for $\phi=3\pi/2$.

\section{Experimental observation of the fractional ac Josephson effect}
\label{sec:experiment}

Conceptually, the setup for experimental observation of the fractional
ac Josephson effect is straightforward.  One should apply a dc voltage
$V$ to the junction and measure frequency spectrum of microwave
radiation from the junction, expecting to detect a peak at the
fractional frequency $eV/\hbar$.  Higher harmonics, such as
$2eV/\hbar$, may also be present because of Eq.\ (\ref{I-phi/2}) and
circuit nonlinearities, but an observation of the 1/2 subharmonic of
the conventional Josephson frequency $2eV/\hbar$ would be the
signature of the effect.

Josephson radiation at the conventional frequency $2eV/\hbar$ was
first observed experimentally almost 40 years ago in Kharkov
\cite{Yanson1,Yanson2}, followed by further work
\cite{Langenberg,Yanson3}.  In Ref.\ \cite{Yanson2}, the spectrum of
microwave radiation from tin junctions was measured, and a sharp peak
at the frequency $2eV/\hbar$ was found.  Without any attempt to match
impedances of the junction and waveguide, Dmitrenko and Yanson
\cite{Yanson2} found the signal several hundred times stronger than
the noise and the ratio of linewidth to the Josephson frequency less
than $10^{-3}$.  More recently, a peak of Josephson radiation was
observed in Ref.\ \cite{Lukens} in indium junctions at the frequency 9
GHz with the width 36 MHz.  In Ref.\ \cite{Mulller}, a peak of
Josephson radiation was observed around 11 GHz with the width 50 MHz
in $\rm Bi_2Sr_2CaCu_2O_8$ single crystals with the current along the
$\bm{c}$ axis perpendicular to the layers.

To observe the fractional ac Josephson effect predicted in this paper,
it is necessary to perform the same experiment with the
$45^\circ/45^\circ$ cuprate junctions shown in Fig.\ 
\ref{fig:d-wave}(a).  For control purposes, it is also desirable to
measure frequency spectrum for the $0^\circ/0^\circ$ junction shown in
Fig.\ \ref{fig:d-wave}(b), where a peak at the frequency $eV/\hbar$
should be minimal.  It should be absent completely in a conventional
$s$-$s$ junction, unless the junction enters a chaotic regime with
period doubling \cite{Miracky,Lobb}.  The high-$T_c$ junctions of the
required geometry can be manufactured using the step-edge technique.
Bicrystal junctions are not appropriate, because the crystal axes
$\bm{a}$ and $\bm{b}$ of the two superconductors are rotated relative
to each other in such junctions.  As shown in Fig.\ 
\ref{fig:d-wave}(a), we need the junction where the crystal axes of
the two superconductors have the same orientation.  Unfortunately,
attempts to manufacture Josephson junctions from the Q1D organic
superconductors $\rm(TMTSF)_2X$ failed thus far.

The most common way of studying the ac Josephson effect is observation
of the Shapiro steps \cite{Shapiro}.  In this setup, the Josephson
junction is irradiated by microwaves with the frequency $\omega$, and
steps in dc current are detected at the dc voltages
$V_n=n\hbar\omega/2e$.  Unfortunately, this method is not very useful
to study the effect that we predict.  Indeed, our results are
effectively obtained by the substitution $2e\to e$.  Thus, we expect
to see the Shapiro steps at the voltages
$V_m=m\hbar\omega/e=2m\hbar\omega/2e$, i.e.\ we expect to see only
\emph{even} Shapiro steps.  However, when both terms are present in
Eq.\ (\ref{I-phi/2}), they produce both even and odd Shapiro steps, so
it would be difficult to differentiate the novel effect from the
conventional Shapiro effect.  Notice also that the so-called
fractional Shapiro steps observed at the voltage
$V_{1/2}=\hbar\omega/4e$ corresponding to $n=1/2$ have nothing to do
with the effect that we propose.  They originate from the higher
harmonics in the current-phase relation $I\propto\sin(2\phi)$.  The
fractional Shapiro steps have been observed in cuprates
\cite{Early,Horng,Borisenko}, but also in conventional $s$-wave
superconductors \cite{Clarke}.  Another method of measuring the
current-phase relation in cuprates was employed in Ref.\ \cite{ZK},
but connection with our theoretical results is not clear at the
moment.

\section{Conclusions}

In this paper, we study suitably oriented $p_x$-$p_x$ or $d$-$d$
Josephson junctions, where the superconductors on both sides of the
junction originally have the surface Andreev midgap states.  In such
junctions, the Josephson current $I$, carried by the hybridized subgap
Andreev bound states, is a $4\pi$-periodic function of the phase
difference $\phi$: $I\propto\sin(\phi/2)$, in agreement with Ref.\ 
\cite{Kitaev}.  Thus, the ac Josephson current should exhibit the
fractional frequency $eV/\hbar$, a half of the conventional Josephson
frequency $2eV/\hbar$.  In the tunneling limit, the Josephson current
is proportional to the first power of the electron tunneling
amplitude, not the square as in the conventional case
\cite{Tanaka2,Riedel,Barash}. Thus, the Josephson current in the
considered case is carried across the interface by single electrons
with charge $e$, rather than by Copper pairs with charge $2e$.  The
fractional ac Josephson effect can be observed experimentally by
measuring frequency spectrum of microwave radiation from the junction
and detecting a peak at $eV/\hbar$.

In $p_x$-$p_x$ junctions with nonparallel orientation of the
spin-polarization vectors $\bm{n}$, the Josephson current depends on
the relative angle between the vectors $\bm{n}$ \cite{Vaccarella}.
The Josephson current is permitted between singlet and triplet
superconductors, but, in the static thermodynamic limit, the
current-phase relation is $\pi$-periodic \cite{Yip}.  The $s$-$p$
junction has two equal minima in energy at $\phi=\pi/2$ and $3\pi/2$
\cite{Asano}, characterized by accumulation of the up or down spins
(oriented relative to the vector $\bm{n}$) in the junction.  In
Josephson junctions between chiral $p_x\pm ip_y$-wave superconductors,
the Andreev bound states are also chiral.  In the static thermodynamic
limit, the current-phase relation has the period of $2\pi$ in the
chiral $p$-$p$ junctions \cite{Barash-chiral} and the period of $\pi$
in the chiral $s$-$p$ junctions \cite{Asano}.

\begin{acknowledgments}
  VMY and HJK thank F.~C.~Wellstood, C.~J.~Lobb, and A.~Yu.~Kitaev for
  useful discussions.  KS thanks S.~M.~Girvin for support.  The work
  was supported by the NSF Grant DMR-0137726.
\end{acknowledgments}

\emph{Note added in proof.} The fractional Josephson effect discussed
in our paper is similar to the fractional quantum Hall effect
\cite{Thouless}.  Both involve existence of several equivalent ground
states, whose energy levels cross: Compare Fig.\ \ref{fig:JJ}c of our
paper and Fig.\ 2a of Ref.\ \cite{Thouless}.


\begin{thebibliography}{00}
\bibitem{2phi} The current-phase relation $I\propto\sin(\phi/2)$ that
  we propose should not be confused with another unconventional
  current-phase relation $I\propto\sin(2\phi)$ with the period $\pi$,
  which was predicted theoretically for junctions between $s$- and
  $d$-wave superconductors \cite{Yip,Tanaka-sd,Zagoskin-sd}, $s$- and
  $p$-wave superconductors \cite{Yip,Tanaka-sp,Asano}, and for the
  $0^\circ/45^\circ$ junctions between two $d$-wave superconductors
  \cite{Wendin-ac}.
\bibitem{A-B} Eq.\ (\ref{eq:I_a}) can be equivalently obtained as a
  difference between the right- and left-moving currents using the
  formula $I_a=ev_F(|A_a|^2-|B_a|^2)$ with the normalization condition
  $\int dx\,|\psi^2|=1$ \cite{Furusaki99}.
\bibitem{double} It is well-known that using the BdG formalism for the
  spin up and down sectors constitutes double-counting \cite{Leggett}.
  Thus, the total energy and current are obtained by adding the
  contributions of the spin up and down sectors and dividing the
  result by two.
\end{thebibliography}
\end{document}